\documentclass[onecollarge,natbib]{svjour2}
\bibpunct{[}{]}{;}{n}{}{,} 
\smartqed  
\usepackage{graphicx}

\usepackage[english]{babel}
\usepackage[utf8x]{inputenc}

\usepackage{latexsym}
\usepackage{amsmath}
\usepackage{amssymb}
\usepackage{amsfonts}
\usepackage{cancel}
\usepackage{bm}

\def\tstrut{\vrule height2.75ex depth0pt width0pt} 
\journalname{Topical Collection: EFB23}
\begin{document}

\title{From $J/\psi$ to LHCb pentaquark
}
\subtitle{Few-body issues on the hadron spectrum}


\author{F. Fern\'andez, D. R. Entem, P. G. Ortega, J. Segovia
}


\institute{
F. Fern\'andez \at
Grupo de F\'isica Nuclear and Instituto Universitario de F\'isica 
Fundamental y Matem\'aticas (IUFFyM), Universidad de Salamanca, E-37008 
Salamanca, Spain \\ \email{fdz@usal.es}           
\and
D. R. Entem \at
Grupo de F\'isica Nuclear and Instituto Universitario de F\'isica Fundamental y 
Matem\'aticas (IUFFyM), Universidad de Salamanca, E-37008 Salamanca, Spain \\
\email{entem@usal.es} 
\and
P. G. Ortega \at
Instituto de F\'isica Corpuscular (IFIC), Centro Mixto CSIC-Universidad de 
Valencia, ES-46071 Valencia, Spain \\
\email{pgortega@ific.uv.es}
\and
J. Segovia \at
Physik-Department, Technische Universit\"at M\"unchen, 
James-Franck-Str.~1, 85748 Garching, Germany \\
\email{jorge.segovia@tum.de}
}

\date{Received: date / Accepted: date}

\maketitle

\begin{abstract}
The original charmonium two-body problem, like the $c\bar c$ structure of the 
$J/\psi$ meson, has become more involved in the last few years with the 
discovery of new resonances such as the tentative molecular state $X(3872)$ or 
the possible pentaquark one $P_{c}(4380)^{+}$. We discuss herein how these 
exotic states (and others) can be described in a unified way adding higher Fock 
state components to the naive quark model picture. In particular, we present our 
theoretical results on the pentaquark states $P_{c}(4380)^{+}$ and 
$P_{c}(4450)^{+}$, and on the new charmonium-like resonances $X(4140)$, 
$X(4274)$, $X(4500)$ and $X(4700)$ that have been reported in $2016$ by the 
LHCb Collaboration.
\end{abstract}


\vspace*{-0.50cm}
\section{Introduction}
\label{sec:intro}

Charmonium history began in $1974$ with the discovery of the so-called $J/\psi$ 
particle. Soon after that, it was realized that $J/\psi$ could be described as 
a bound state of a charm quark and a charm antiquark, namely as a two-body 
problem. Moreover, owing to the large mass of the charm quark, the system could 
be considered in a first place by its nonrelativistic nature and thus solving 
the Schr\"odinger equation with a well motivated QCD potential was a successful 
approach to the problem at hand. 

The discovery in $2003$ of the $X(3872)$ resonance made evident that the naive 
quark model picture is not enough to describe its properties. For instance, the 
large mass of the $X(3872)$ and its zero isospin value point out that it is 
made by charm quarks; on the other hand, the isospin violating decay 
$X(3872)\to J/\psi \rho$ makes the $X(3872)$ difficult to accommodate as 
charm-anticharm state. However, the decay mentioned above can be naturally 
explained in a picture in which the $X(3872)$ is understood as a $DD^{\ast}$ 
bound state, namely a four-quark bound-state problem.

During the last decade, experimental observations have revealed the existence 
of a large number of unexpected states such as the $X(3872)$, culminating 
recently with the observation by the LHCb collaboration in $2015$ of two 
charmonium pentaquark states $P_{c}(4380)^{+}$ and 
$P_{c}(4450)^{+}$~\cite{Aaij:2015tga}. Therefore, we have gone from a two-body 
problem to a five-body problem passing trough a four-body one in less than 15 
years.

In this proceedings contribution we present a unified description of all 
these states within the framework of a constituent model of quarks. The model 
has already been applied to two- and four-quark structures and is now 
being applied to the study of five-quark states. In addition, we discuss herein 
a set of four resonances recently measured in the LHCb~\cite{Aaij:2016iza}: 
$X(4140)$, $X(4274)$, $X(4500)$ and $X(4700)$, which have been considered as 
new multiquark states~\cite{Chen2016} but in our scheme most of them appear as 
simple quark-antiquark structures.


\vspace*{-0.50cm}
\section{The constituent quark model}
\label{sec:model}

Our constituent quark model (CQM) is based on the assumption that the light 
massless light quark acquires a dynamical, momentum dependent mass namely the 
constituent quark mass as a consequence of the dynamical chiral symmetry 
breaking of the QCD Lagrangian at some momentum scale. This constituent quark 
mass is frozen at low momenta at a value around $300\,{\rm MeV}$ and explains 
the $98\%$ of the mass of the proton. The simplest Lagrangian which mimics this 
situation must contain chiral fields to compensate the constituent mass term. 
These chiral fields induce boson exchanges between quarks. In the heavy quark 
sector, chiral symmetry is explicitly broken and we do not need additional 
fields. However, the chiral fields introduced above provide a natural way to 
incorporate one-pion (OPE) interactions in the molecular (four-quark) dynamics. 
Higher pion exchanges than OPE are effectively included by scalar boson 
exchanges.

Besides the dynamically broken chiral symmetry, the other two main properties 
of QCD that are incorporated in our quark model are confinement and asymptotic 
freedom. To mimic such features we use a linear screened confining potential 
and a perturbative one-gluon exchange interaction. The detailed explanation and 
parametrization of all these interactions can be found in 
Ref.~\cite{Vijande:2004he}, updated in Ref.~\cite{Segovia:2008zz}.

To obtain the solutions in the two-body case, we solve the Schr\"odinger 
equation using the Gaussian Expansion method~\cite{Hiyama:2003cu} whereas the 
four and five-body problem are solved using the Resonating Group 
Method~\cite{Wheeler:1937zz}. Two-quark and multiquark configurations are 
coupled using the $^3P_0$ method~\cite{LeYaouanc:1972ae}. Details of the 
calculations are given in Refs.~\cite{Ortega:2010qq, Ortega:2016mms}


\begin{table}[!t]
\caption{Masses and width of the different pentaquark states}
\centering
\label{t1} 
\begin{tabular}{cccccc}
\hline\noalign{\smallskip}
Molecule & $J^P$ & $I$ & $M$ (MeV) & $\Gamma(J/\psi p)$ (MeV) & $\Gamma(\bar 
D^{\ast}\Lambda_{c})$ (MeV) \\
\tableheadseprule\noalign{\smallskip}
$\bar D\Sigma_{c}^{\ast}$ & $\frac{3}{2}^-$ & $\frac{1}{2}$ & $4385.0$ & $10.0$ 
& $14.7$ \\[0.5ex]
\hline
\tstrut
$\bar D^{\ast}\Sigma_{c}$ & $\frac{1}{2}^-$ & $\frac{1}{2}$ & $4458.9$ & $5.3$ 
& $63.6$ \\
$\bar D^{\ast}\Sigma_{c}$ & $\frac{3}{2}^-$ & $\frac{1}{2}$ & $4461.3$ & $0.8$ 
& $21.2$ \\
$\bar D^{\ast}\Sigma_{c}$ & $\frac{3}{2}^+$ & $\frac{1}{2}$ & $4462.7$ & $0.2$ 
& $6.3$ \\
\noalign{\smallskip}\hline
\end{tabular}
\end{table}

\vspace*{-0.50cm}
\section{The $P_{c}(4380)^{+}$ and $P_{c}(4450)^{+}$ resonances}
\label{sec:pentaquarks}

We consider the $\bar{D}\Sigma_{c}^{\ast}$ and $\bar{D}^{\ast}\Sigma_{c}$ 
thresholds which are the only ones close to the mass of the $P_{c}(4380)^{+}$ 
and $P_{c}(4450)^{+}$ resonances and also where a sizable residual interaction 
mainly due to pion exchanges can be expected.

One can see in Table~\ref{t1} that in the mass region of the $P_{c}(4380)^{+}$ 
we obtain a $\bar{D}\Sigma_{c}^{\ast}$ bound state with 
$J^{P}=\frac{3}{2}^{-}$. Its mass is very close to the experimental one and 
should, in principle, be identified with the observed $P_{c}(4380)^{+}$ 
structure.

Referring to the channel $\bar{D}^{\ast}\Sigma_{c}$, we find three almost 
degenerated states with a mass around $4460\,{\rm MeV}$ that makes them natural 
candidates for the $P_{c}(4450)^{+}$ resonance. The total spin and parity of 
these states are $J^{P}=\frac{1}{2}^{-}$, $\frac{3}{2}^{-}$ and 
$\frac{3}{2}^{+}$. Their degeneration may be the origin of the uncertainty in 
the experimental determination of the total spin and parity for the 
$P_{c}(4450)^{+}$ resonance.

It is remarkable that the three bound states predicted by our model decay into 
$\bar{D}^{\ast}\Lambda_{c}$ final state with a partial decay width which is 
generally equal to or larger than the width corresponding to the $J/\psi p$ 
decay channel. This suggests that the $\bar D^{\ast}\Lambda_{c}$ final state is 
a suitable decay channel for studying the properties of these resonances. In 
particular, the width of the predicted $\bar D^{\ast}\Sigma_{c}$ resonance with 
$J^{P}=\frac{1}{2}^{-}$ is twelve times larger through the $\bar 
D^{\ast}\Lambda_{c}$ channel than through the $J/\psi p$ channel, making this 
decay an important check of our prediction.

Concerning the parity of the states, a molecular scenario is not the most 
convenient to obtain positive parity states because, being the $\bar D^{(*)}$ 
mesons and the $\Sigma_c^{(*)}$ baryons of opposite parity, the relative 
angular momentum should be at least $L=1$ (P-wave) which will be above S-waves.
This is reflected in the fact that the states with positive parity in 
Table~\ref{t1} are those with smaller binding energies.


\begin{table}[!t]
\caption{Naive quark-antiquark spectrum in the region of 
interest of the LHCb~\cite{Aaij:2016iza, Aaij:2016nsc} for the $0^{++}$ and 
$1^{++}$ channels.}
\centering
\label{tab:qqbar} 
\begin{tabular}{ccccc}
\hline\noalign{\smallskip}
State & $J^{PC}$ & $nL$  & Theory (MeV) & Experiment (MeV) \\
\tableheadseprule\noalign{\smallskip}
$\chi_{c0}$ & $0^{++}$ &  $3P$ & $4241.7$  & $-$\\
            &          &  $4P$ & $4497.2$  & $4506\pm 11^{+12}_{-15}$ \\
            &          &  $5P$ & $4697.6$  & $4704\pm 10^{+14}_{-24}$ \\
\hline
$\chi_{c1}$ & $1^{++}$ &  $3P$ & $4271.5$  &  $4273.3\pm 8.3$ \\
            &          &  $4P$ & $4520.8$  &      $-$            \\
            &          &  $5P$ & $4716.4$  &     $-$             \\           
\noalign{\smallskip}\hline
\end{tabular}
\end{table}

\begin{table}[!t]
\caption{Mass, total width (in MeV), and $c\bar c$ component 
probabilities (in \%) for the $X(4274)$ meson, obtained from the coupled 
channel calculation described in the text}
\centering
\label{tab:r3}
\begin{tabular}{cccccc}
\hline\noalign{\smallskip}
Mass & Width & ${\cal P}_{c\bar c}$ & ${\cal P}_{D_{s}D_{s}^{\ast}}$ & ${\cal 
P}_{D_{s}^{\ast}D_{s}^{\ast}}$ & ${\cal P}_{J/\psi\phi}$ \\
\tableheadseprule\noalign{\smallskip}
$4242.4$ & $25.9$ & $48.7$ & $43.5$ & $5.0$ & $2.7$ \\
\noalign{\smallskip}\hline
\end{tabular}
\end{table}

\begin{table}[!t]
\caption{Probabilities, in \%, of $nP$ $c\bar c$ components in the total wave 
function of the $X(4274)$ meson.}
\centering
\label{tab:r4}  
\begin{tabular}{ccccccc}
\hline\noalign{\smallskip}
Mass (MeV) & ${\cal P}_{c\bar c}$ & ${\cal P}_{1P}$ & ${\cal P}_{2P}$ & 
${\cal P}_{3P}$ & ${\cal P}_{4P}$ & ${\cal P}_{(n>4)P}$ \\
\tableheadseprule\noalign{\smallskip}
$4242.4$ & $48.7$& $0.000$ & $0.370$  & $99.037$ & $0.488$ & $0.105$ \\
\noalign{\smallskip}\hline
\end{tabular}
\end{table}

\vspace*{-0.50cm}
\section{The $X(4140)$, $X(4274)$, $X(4500)$ and $X(4700)$ resonances}
\label{sec:LHCbresonances}

Table~\ref{tab:qqbar} shows the calculated naive quark-antiquark spectrum in 
the region of interest of the LHCb for the $J^{PC}=0^{++}$ and $1^{++}$ 
channels. A tentative assignment of the theoretical states with the 
experimentally observed mesons at the LHCb experiment is also given. It can be 
seen that the naive quark model is able to reproduce all the new LHCb 
resonances except the $X(4140)$. The $X(4274)$, $X(4500)$ and $X(4700)$ appear 
as conventional charmonium states with quantum numbers $3^{3}P_{1}$, 
$4^{3}P_{0}$ and $5^{3}P_{0}$, respectively. A complete study of the decay 
widths of these states has been performed in Ref.~\cite{Ortega:2016hde} showing 
that the total decay width of the $X(4274)$, $X(4500)$ and $X(4700)$ are, 
withing errors, in reasonable agreement with the data.

To gain some insight into the nature of the $X(4140)$, that does not appear as 
a quark-antiquark state, and to see how the coupling with the open-flavour 
thresholds can modify the properties of the naive quark-antiquark states 
predicted above, we have performed a coupled-channel calculation including the 
$D^{\ast}D_{1}^{(\prime)}$, $D_{s}D_{s}$, $D_{s}^{\ast}D_{s}^{\ast}$ and 
$J/\psi\phi$ channels for the $J^{PC}=0^{++}$ sector; and the 
$D_{s}D_{s}^{\ast}$, $D_{s}^{\ast}D_{s}^{\ast}$ and $J/\psi\phi$ ones for the 
$J^{PC}=1^{++}$ sector. These are the allowed channels whose thresholds are in 
the region studied by the LHCb.

Our results for the $J^{PC}=0^{++}$ channel can be found in 
Ref.~\cite{Ortega:2016hde}. Therein, we conclude first that the net effect of 
coupling the thresholds to both naive quark-antiquark states is to modify the 
mass of the bare $c\bar c$ states in a modest amount. The second observation is 
that the total decay widths are significantly reduced.

For the coupled-channel calculation in the $J^{PC}=1^{++}$ channel, we found 
only one state with mass $4242.4\,{\rm MeV}$ and total decay width $25.9\,{\rm 
MeV}$. This state is made by $48.7\%$ of the $3P$ charmonium state and by 
$43.5\%$ of the $D_{s}D_{s}^{\ast}$ component. Our theoretical results are 
given in Tables~\ref{tab:r3} and~\ref{tab:r4}.

\begin{figure}[!t]
\begin{center}
\includegraphics[width=0.40\textwidth,angle=-90]{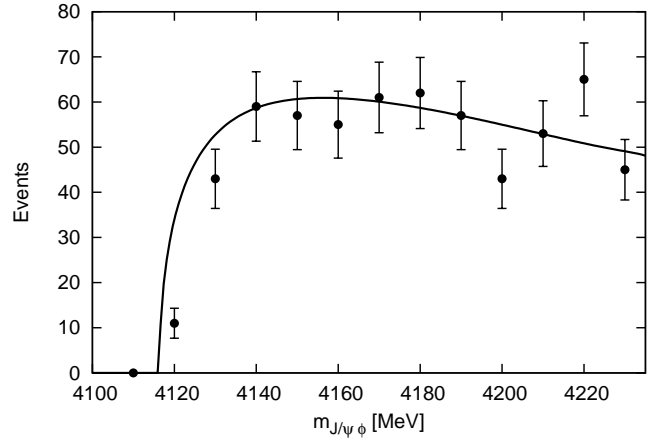}
\caption{\label{fig:f2} Line-shape prediction of the $J/\psi\phi$ channel. The 
curve shows the production of $J/\psi\phi$ pairs via direct generation from a 
point-like source plus the production via intermediate $c\bar c$ states. Note 
that the production constant has been fitted to the data (see 
Ref~\cite{Ortega:2016hde} for the details).}
\end{center}
\end{figure}

As we do not find any signal for the $X(4140)$, neither bound nor virtual, we 
analyze the line shape of the $J/\psi\phi$ channel as an attempt to explain the 
$X(4140)$ as a simple threshold cusp (see again Ref.~\cite{Ortega:2016hde} for 
details). Figure~\ref{fig:f2} compares our result with that reported by the LHCb 
Collaboration in the $B^+\to J/\psi\phi K^+$ decays. The rapid increase observed 
in the data near the $J/\psi\phi$ threshold corresponds to a bump in the 
theoretical result just above such threshold. This cusp is too wide to be 
produced by a bound or virtual state below the $J/\psi\phi$ threshold. 

\vspace*{-0.50cm}
\section{Summary}

As a summary, our results confirm the fact that there are several states with a 
$\bar{D}^{(\ast)}\Sigma_{c}^{(\ast)}$ structure in the vicinity of the masses 
of the $P_{c}(4380)^{+}$ and $P_{c}(4450)^{+}$ pentaquarks reported by the LHCb.

Concerning the other resonances, three of them, namely $X(4274)$, $X(4500)$ and 
$X(4700)$, are consistent with bare quark-antiquark states with quantum numbers 
$J^{PC}=1^{++} (3P)$, $J^{PC}=0^{++} (4P)$ and $J^{PC}=0^{++} (5P)$, 
respectively.

In the $1^{++}$ sector, we do not find any pole in the mass region of the 
$X(4140)$. However, the scattering amplitude shows a bump just above the 
$J/\psi\phi$ threshold which reproduces the fast increase of the experimental 
data. Therefore, the structure showed by this data around $4140\,{\rm MeV}$ 
should be interpreted as a cusp due to the presence of the $D_{s}D_{s}^{\ast}$ 
threshold.


\vspace*{0.25cm}
{\bf Acknowledgments} This work has been partially funded by Ministerio de 
Ciencia y Tecnolog\'\i a under Contract no. FPA2013-47443-C2-2-P, by the 
Spanish Excellence Network on Hadronic Physics FIS2014-57026-REDT, and by the 
Junta de Castilla y Le\'on under Contract no. SA041U16. P. G. Ortega 
acknowledges the financial support of the Spanish Ministerio de Econom\'ia y 
Competitividad and European FEDER funds under the contracts 
FIS2014-51948-C2-1-P. J. Segovia acknowledges the financial support from 
Alexander von Humboldt Foundation.



\begin{thebibliography}{13}

\bibitem{Aaij:2015tga} 
R.~Aaij {\it et al.} [LHCb Collaboration],
(2015) Phys.\ Rev.\ Lett.\  {\bf 115}, 072001.

\bibitem{Aaij:2016iza} 
R.~Aaij {\it et al.} [LHCb Collaboration],
(2016) arXiv:1606.07895 [hep-ex].

\bibitem{Chen2016} 
H.~X.~Chen, W.~Chen, X.~Liu and S.~L.~Zhu,
(2016) Phys.\ Rept.\  {\bf 639}, 1.

\bibitem{Vijande:2004he} 
J.~Vijande, F.~Fernandez and A.~Valcarce,
(2005) J.\ Phys.\ G {\bf 31}, 481.
  
\bibitem{Segovia:2008zz} 
J.~Segovia, A.~M.~Yasser, D.~R.~Entem and F.~Fernandez,
(2008) Phys.\ Rev.\ D {\bf 78}, 114033.
  
\bibitem{Hiyama:2003cu} 
E.~Hiyama, Y.~Kino and M.~Kamimura,
(2003) Prog.\ Part.\ Nucl.\ Phys.\  {\bf 51}, 223.

\bibitem{Wheeler:1937zz} 
J.~A.~Wheeler,
(1937) Phys. Rev. {\bf 52}, 223.

\bibitem{LeYaouanc:1972ae}
A.~Le Yaouanc, L.~Oliver, O.~Pene and J.C.~Raynal,
(1973) Phys. Rev. {\bf D8}, 2223-2234.

\bibitem{Ortega:2010qq} 
P.~G.~Ortega, J.~Segovia, D.~R.~Entem and F.~Fernandez,
(2010) Phys.\ Rev.\ D {\bf 81}, 054023.
  
\bibitem{Ortega:2016mms} 
P.~G.~Ortega, J.~Segovia, D.~R.~Entem and F.~Fernandez,
(2016) Phys.\ Rev.\ D {\bf 94}, no. 7, 074037.

\bibitem{Ortega:2016hde} 
P.~G.~Ortega, J.~Segovia, D.~R.~Entem and F.~Fernández,
(2016) arXiv:1608.01325 [hep-ph]
 
\end{thebibliography}

\end{document}